 \documentclass[final,1p,times,authoryear]{elsarticle}

\usepackage{graphicx}
\usepackage{amssymb}
\usepackage{color}

\usepackage{lineno}
\usepackage{natbib}

\biboptions{}

\journal{New Astronomy Reviews}

\begin{document}

\begin{frontmatter}


\title{Dark matter halo concentrations: a short review}



\author[1,2]{Chiamaka Okoli}
\ead{c2okoli@uwaterloo.ca}

\address[1]{Perimeter Institute for Theoretical Physics\\
31 Caroline Street North\\
Waterloo, ON, Canada N2L 2Y5 }
\address[2]{University of Waterloo\\
200 University Avenue West\\
Waterloo, ON, Canada N2L 3G1}

\begin{abstract}
This review analyzes the state and advancement of the dark matter halo concentrations over the last two decades.  It begins with presenting the article that brought the field to the limelight and then follows through with other research works that studied the concentrations of dark matter haloes over the ages. Besides the discussion of the halo mass-concentration relation and its evolution, we examine the effects of cosmology, subhaloes and environment on the relation. In addition to theoretical halo concentrations, observational dark matter halo concentrations are also considered. This review synthesizes the progress in this field into a clear piece of article.
\end{abstract}

\begin{keyword}
Dark Matter  \sep Halos \sep Concentration


\end{keyword}

\end{frontmatter}
\section{Introduction to halo concentrations}
In this article, we will present a chronological history of the dark matter halo concentrations and their relation to mass primarily, occasionally relating the concentration to other halo parameters, and review the state of the halo mass-concentration relation to present. This work focuses on a lot of literature and articles related to the halo mass and concentration from present and dating back to the presentation of the density profiles of dark matter haloes by \citet{1996ApJ...462..563N}. 

From N-body simulations of dark matter particles, the density profile of dark matter haloes have been shown to follow the cuspy two-parameter profile given below:
\begin{equation}
\rho_{\rm NFW} = \frac{\rho_c \delta_c}{\left(\frac{r}{r_s}\right)\left(1 + \frac{r}{r_s}\right)^2},
\end{equation}
where $\rho_c$ is the critical density of the universe, $r_s$ is a scale radius where the density profile is isothermal and $\delta_c$ is an overdensity parameter. This form of the density profile (hereafter the NFW density profile) is seen to be universal for different masses and variants of the universe such as flat, open, and closed universes. From the form of the profile, $\rho \propto 1/r$ in the innermost regions of the halo and $ \propto 1/r^3$ in the outer regions of the halo. The above density profile may be rewritten in terms of the halo mass and halo concentration, with the concentration defined as $c_{200} \equiv \frac{r_{200}}{r_s}$. Another variant of the concentration that may be found in the literature is $c_{vir} \equiv \frac{r_{vir}}{r_s}.$ Due to the fact that the latter definition of the halo concentration varies a lot more with the chosen cosmology and redshift, we shall focus this review on $c_{200}$, which gives the most universal relation \citep{2015ApJ...799..108D}.  We shall discuss $c_{vir}$ where necessary, not excluding in its entirety from our discussion. From N-body simulations, there exists a relation between the halo mass and halo concentration,  initially found by \citet{1996ApJ...462..563N} and discussed a lot by other articles in this review -- small mass haloes have higher concentrations while the large mass haloes have lower concentrations. This relation has, over the years, been studied and analyzed by a number of authors. Analytic arguments related to the mass-concentration relation have also been presented by several other authors; these works of research will be reviewed in different sections of this article. 

Although the NFW shows traits of universality across all scales -- from galaxies to clusters -- and cosmology, this was not expected due to the variation of the effective slope of the power spectrum across these scales (from $n_{\rm eff} \approx -3$ to $n_{\rm eff} \approx -2$ respectively). \citet{Navarro:1996gj} expected that the galaxy-scale haloes should show shallower density profiles while the cluster scale haloes display steeper density since larger n's are expected to have steeper profiles in an open universe (\cite{1985ApJ...297...16H}). However, results from simulations revealed that the slopes of the density profiles were steeper for low mass haloes, at fixed fraction of the viral radius and shallower for the cluster scale haloes. Given that steeper profiles were observed in simulations for haloes of a given mass for larger spectral indices $n$, the correlation between the concentration and mass may actually be controlled by the spectral index.

In addition to the NFW profile, the Einasto profile \citep{2004MNRAS.349.1039N} was proposed as a better fit to the density profiles of dwarfs, galaxies and clusters from N-body simulations. The idea being that the logarithmic slope of the density profile at highly resolved radii were observed to be shallower with decreasing radii rather than converging to a given value (say -1 for the NFW profile). The form of the Einasto profile is given by
\begin{equation}
\rho_{\rm Einasto} = \rho_{-2} \exp{\left\lbrace\frac{-2}{\alpha}\left[\left(\frac{r}{r_{-2}}\right)^{\alpha} -1\right]\right\rbrace}.
\end{equation}
The parameters $\rho_{-2}$ and $r_{-2}$ are the density and the radius respectively at the scale radius (the radius where the logarithmic slope of the density profile is isothermal, -2). In this case, the concentration is also defined as $c_{200} \equiv \frac{r_{200}}{r_{-2}}$. The shape parameter, $\alpha$, parameter determines the shape of the density profile and how slow or fast the slope changes with radius. In addition, this parameter tailors the fit to every individual halo and has been found to show a mass dependence. In terms of the linearly extrapolated overdensity $\delta_c$ and the variance in density fluctuations $\sigma(M,z)$, the variation of the shape parameter in mass is parametrized with the peak height parameter $\nu \equiv \frac{\delta_c}{\sigma(M,z)}$:
\begin{eqnarray}
\alpha (\nu) &=& 0.155 + (0.0095 \nu^2)\nonumber \\
\alpha (\nu) &=& 0.115 + (0.014 \nu^2)
\end{eqnarray}
for \citet{2008MNRAS.387..536G} and \citet{2014arXiv1411.4001K} respectively. Although this review focuses on the concentrations derived from the NFW profile, we will occasionally discuss the concentrations from the Einasto profile. We will attempt to present this review based on most articles related to halo mass and concentration. However, we do not claim to have exhausted all the articles written in this field. 

\section{Mass-concentration relation}
Halo concentrations are usually a decreasing function of mass, albeit with some characteristic feature. In the article by NFW, the authors attributed the mass-concentration relation to the density of the universe at the time of the collapse of the halo. Adopting a definition for the formation time in terms of a given fraction of the final mass, $f \ll 1$, different haloes were assigned formation times based on the Press-Schechter formalism. Thereafter the characteristic density, $\delta_c$ was scaled to the density of the universe at the time of formation of the halo. The correlation between mass and density or mass and concentration were then made for various values of $f.$

This assertion have been corroborated by  a number of authors who link the halo concentrations to their mass accretion histories (MAH). The MAH is the increase in mass of the main progenitor in a halo. The average MAH of the main progenitor is usually fit to a function that depends on the formation epoch of the halo. The result is a MAH with individual shapes dependent on the mass of the haloes. The massive haloes tend to accumulate mass till late times while the small mass haloes asymptote earlier. In order words, the small mass haloes assemble most of their masses earlier while the large mass haloes witness slower mass accretion and assemble the majority of their mass at later times. Thus, haloes undergoing recent mergers tend to have latter formation times and lower concentrations. The mass accretion history of haloes can be described by two distinguishing phases -- and early fast phase and a late slow phase \citep{2003MNRAS.339...12Z}. The MAH is may be fit to a function that depends on the formation epoch of the halo and is independent of the merging phase of the halo. It is important to note that the concentration of a halo is fixed by its mass and is independent of the the time of observation \citep{0004-637X-568-1-52}. Given the formation epoch for a halo of a given mass, one may then estimate the concentration using an empirical relation between the concentration and the formation epoch. 

Given a definition of formation time, the distribution of formation redshifts can be estimated for a range of masses. This distribution is a log-normal distribution, whose mean decreases as the mass increases. In another prescription, formation times may be estimated from halo merger trees and then related to halo concentrations. The merger tree is the increase in mass of the main halo. Merger trees may be measured from N-body simulations or predicted from the extended Press-Schechter (EPS) formalism. Formation times from the EPS formalism have been observed to be about $25\%$ larger than those that are being measured from simulations \citep{0004-637X-568-1-52}. 

As pointed out by \citet{Navarro:1996gj}, the correlation between the concentration and mass of haloes may be explained by the spectral index on the given mass scale. For a given mass and fixed matter density $\Omega_{m0}$, haloes with higher values of $n$ were observed to be more concentrated. Likewise, haloes formed in a universe with lower $\Omega_0$ are less concentrated than those with higher matter density. Even though haloes collapse earlier in a low $\Omega_{m0}$ universe, the change in $\Omega_{m0}$ outweighs the change in the collapse redshift since the collapse density is related to both by 
\begin{equation}
\delta_c = \Omega_{m0}(1 + z_{col})^3
\end{equation}

In summary, halo concentrations are fixed by the formation epoch of a halo, albeit dependent on some of the cosmological parameters such as $n$ and $\Omega_{m0}$.

\section{Theoretical dark matter halo mass - concentration relations}
\subsection{N-body simulations overview}

In this section, we present a number of theoretical halo concentrations and discuss the physical explanations for the relation between dark matter mass and concentrations. Most of these studies are based on results from N-body dark matter simulations. This involves generating initial conditions similar to that expected from the early universe using a set of codes such as GRAFIC\footnote{http://www.projet-horizon.fr/article258.html}, 2LPTIC\footnote{http://cosmo.nyu.edu/roman/2LPT/}, MUSIC \citep{2011MNRAS.415.2101H} and running the particles through gravitational interaction with a set of codes such as GADGET-2 \citep{2005MNRAS.364.1105S} AND PKDGRAV\footnote{http://hpcc.astro.washington.edu/faculty/trq/brandon/pkdgrav.html}. Thereafter, haloes are identified using the spherical overdensity (SO) method -- finding the radius at which the mean halo density is $\Delta_c$ times the critical or the mean density of the universe --  or the friends-of-friends (FOF) method -- connecting particles closer than a fraction of the mean inter particle spacing (usually called linking length), typically assumed to be 0.2. A third type of halo finder similar to the SO (Bound Density Maxima (BDM)) exists and is used in \citet{0004-637X-568-1-52}. For a given set of haloes, most of the results presented below are based on the relaxed subset of the haloes. Relaxed haloes are selected based on a number of criteria which varies according to the authors. These include: the mass fraction of substructure in a given halo \citep{2007MNRAS.381.1450N,2014MNRAS.441..378L}, the offset between the centre of mass and the potential minimum \citep{2007MNRAS.381.1450N,2011MNRAS.411..584M,2014MNRAS.441..378L, 2014arXiv1411.4001K}, the rms  to the fit of the density profile \citep{2007MNRAS.378...55M, 2008MNRAS.391.1940M, 2011MNRAS.411..584M} (NFW or Einasto), the virial ratio (the ratio of twice the kinetic energy to the potential energy) of the  halo \citep{2007MNRAS.381.1450N,2014MNRAS.441..378L,2014arXiv1411.4001K} and the spin parameter \citep{2014arXiv1411.4001K}. These relaxed haloes typically have higher concentrations and smaller scatter for a given mass relative to the unrelaxed haloes, on average \citep{2007MNRAS.381.1450N}. The scatter of the halo concentrations at a given mass will be discussed in a separate section, specifically section \ref{sec:scatter}. Additional selection cuts may be made by restricting to haloes with a given minimum number of particles; for example, 500 particles in \citet{2008MNRAS.391.1940M}. 

Irrespective of the criteria for making the cut on relaxed haloes, one may argue regarding the proper criteria for selecting relaxed haloes (see \citet{2014arXiv1411.4001K} for more details.) Especially on using the virial parameter as a criteria since this relation is usually modified by the effect of the surface pressure to the energy contribution and the external potential energy contribution as haloes are not completely isolated objects. 

Given the identification of haloes in an N-body simulation, the selection of relaxed haloes, halo concentrations are then measured from the fit parameters to the density profile. A mass-concentration relations may then be found. The following subsection looks at some of the various relations in the literature.

\subsection{Theoretical halo mass-concentration relations}
Although halo concentrations were initially thought to be a continually decreasing function of mass, \citet{2003ApJ...597L...9Z} were the first authors to find that the mass-concentration relation flattens at the high mass end at $z=0,$ (verified by a number of authors such as \citet{2007MNRAS.381.1450N,2016MNRAS.456.3068O}) with a minimum concentration of about 3.5. The exact mass at which the relation starts flattening is dependent on the redshift, it decreases to lower masses as the redshift increases. Additionally, the mass-concentration relation becomes weaker at higher redshifts i.e the mass-concentration relation evolves faster at lower redshift ($<3$) than at higher redshift ($>3$) \citep{2007MNRAS.381.1450N,2008MNRAS.390L..64D,2011MNRAS.411..584M,0004-637X-568-1-52,2016MNRAS.456.3068O}. On the mass scales, low-mass haloes evolve more with redshift than high-mass haloes. This correlation between concentration and mass may be explained using the mass accretion rate -- the concentration is constant at the high mass end -- exists because the massive haloes are in the fast accretion phase (where mostly major mergers occur) where the scale radius changes as the viral radius changes and the small mass haloes have higher concentrations because they are in the slow accretion phase (with mainly minor mergers) where the scale radius is more or less constant and the virial radius gradually builds up. 

To predict the mass-concentration relation for any cosmology, redshift or any form of the power spectrum of matter fluctuations, concentrations may be related to a universal model of the mass accretion history (MAH) established through the mass accretion rate \citep{0004-637X-707-1-354}. This model of concentration in terms of the MAH was the first attempt at making halo concentrations universal and also points out to more evidence regarding the redshift evolution of halo concentrations. The results from this model indicate that the haloes evolve in redshift not simply as $\frac{1}{1+z}$ but in more complex form. Thus, a model for the halo concentrations may be built by relating the concentration to the time when the main progenitor had accreted $4\%$ of its final mass (extracted from the MAH). According to \citet{0004-637X-707-1-354}, the time evolution of the concentration is given by 
\begin{equation}
c = 4 \left\lbrace1 + \left(\frac{t}{3.75 t_{0.04}}\right)^{8.4}\right\rbrace^{1/8},
\end{equation}
where $t_{0.04}$ is the time when the main progenitor has accreted $4\%$ of its final mass. In a similar vein, the distribution of halo concentrations may be extracted from the distribution of the formation times of the halo for a given mass. This approach was investigated by \citet{2012MNRAS.422..185G} by characterizing a relation between the distribution of halo formation times and the mass of the main progenitor using simulations. Defining the formation time as the earliest time when the main progenitor of the halo has a mass $m > fM_0$, a strong correlation was found between the halo concentrations and the time provided the time when the halo had assembled about $4\%$ and half of its total mass is known. Given a MAH, one can then cast the concentration-time relation into a concentration-mass relation.

Although mass-concentration relations have been all expressed as a power law in terms of mass up to this point, \citet{2012MNRAS.423.3018P} introduced a more closer universal mass-concentration relation  by expressing the halo concentrations in terms of the root mean square (RMS) of the matter density fluctuations $\sigma (M,z).$ This expression, although not entirely universal in redshift and cosmology \citep{2012MNRAS.423.3018P,2015ApJ...799..108D}, is closer to being universal for a given parameter fit. In addition, the concentrations were measured using a more profile independent method ($\frac{V_{max}}{V_{200}}$) rather than making a given fit to an assumed density profile (say NFW or Einasto). Over six orders of magnitude in mass and and redshifts that range from $0<z<10,$ the results of the concentrations show interesting features. The form of the halo concentrations as a function of mass show three distinct features - a region of decreasing halo concentration with mass (in the low-mass end), a region of flattening, and a region in which the halo concentrations increase with mass (the high-mass end), which was first observed in \citet{2011ApJ...740..102K}. This upturn in the halo mass-concentration relation at the high-mass end is due to the massive haloes having particles falling into mostly radial orbits. To confirm that the upturn in halo concentrations isn't due to non-equilibrium effects, a sample of relaxed haloes displayed similar features. On a different note, \citet{2012MNRAS.427.1322L} do not find the upturn at large masses when considering only relaxed haloes and the similar concentration measurements of \citet{2012MNRAS.423.3018P}. They expressed concerns that the criteria by \citet{2012MNRAS.423.3018P} for selecting the relaxed haloes was less stricter than those of \citet{2012MNRAS.427.1322L}. The explanation for the supposed increase in concentration at the high mass end is that most of the systems are not yet virialized and that the accreted material are most likely experiencing their first pericentric passage. Intuitively, it is not clear that the concentrations increase at the high mass end since halo concentrations are related to the formation time of the haloes. and the high-mass systems have only being recently formed. A  universal relation for concentrations was finally presented in \citet{2015ApJ...799..108D} which is parametrized in terms of the peak height parameter and the local slope of the power spectrum $n_{\rm eff}.$ Expressing halo concentrations in terms of the slope of the power spectrum affects both the normalization and the slope of the concentration-mass relation.

Still on the reason for the occurence of the upturn in halo concentrations, \citet{2013arXiv1303.6158M} made efforts to reconcile the difference in theoretical concentrations from N-body simulations by investigating the different concentration measurements -- using the circular velocity method and using fits to the density profile -- and the binning procedure -- binning in mass versus binning in circular velocity. Both method agree on galaxy scales but differ only on cluster scales. This points out to the fact that the density profiles on the massive objects are less likely to be well fitted by the NFW profile due to the fact that they are most likely to be far from equilibrium. The radial range used in fitting the density profile gives variant concentrations, which was corroborated by \citet{2016MNRAS.459.2106A} for both concentrations and the Einasto shape parameter. \citet{2016MNRAS.459.2106A} also find that the concentration and the shape parameter depend on the minimum number of particles included in making the fit. It is important to note that \citet{2014MNRAS.441.3359D} find that the concentrations using both methods are roughly consistent at $z=0$ and at all masses. The concentrations from the circular velocity method become quite higher at higher redshifts. 

To wrap up the case regarding the upturn in halo concentrations, \citet{2014arXiv1411.4001K} presented a study of halo concentrations from $z=0$ up to $z=5.5.$  In addition to the two features of the concentration-mass relation -- decreasing concentration with mass and a region of flat concentration --  identified by \citet{2003ApJ...597L...9Z}, haloes go through an initial stage of increasing concentration with mass. This feature is more visible in massive haloes -- which are the high-peaked haloes at low redshift. In fact, these massive haloes, better approximated by the Einasto profile with high $\alpha$, are not good fits to the NFW profile. Thus, the high value of the shape parameter affects the concentration of the massive haloes, especially when they are fitted to the NFW profile. The high-$\alpha$ haloes were found to be dominated by high infall velocities at large radii ($r \gg r_{vir}.$), and nearly zero infall velocity in the centre regions. The lack of upturn in the low mass region is attributed to the small value of their shape parameter and the fact that they are good fits of the NW profile.

Other works that relate the halo concentrations to the MAH include \citet{2011MNRAS.411..584M} and \citet{2014MNRAS.441..378L}, that presents a relation between the concentration of a halo fitted to the NFW ($c_{\rm NFW}$) profile and the concentration of the MAH ($c_{\rm MAH}$) fitted to the NFW profile. This is given by 
\begin{equation}
c_{\rm NFW} = 2.9 (1 + 0.614 c_{\rm MAH})^{0.995}.
\end{equation}
This form of the concentration predicts a minimum concentration of $\sim 3.$ \citet{2015MNRAS.452.1217C} also relates halo concentrations to the MAH by measuring the formation redshift of the halo. 

Halo concentrations have also been related to the spectral index of matter density fluctuations \citep{2001ApJ...554..114E}. In this case, the halo concentrations are modelled from the amplitude and shape of the power spectrum of matter density fluctuations using a single parameter. Results from simulations revealed that the halo concentrations in a $\Lambda$CDM universe increase with an increase in the normalization of the power spectrum, $\sigma_8.$ Physically, the redshift for collapse increases with the normalization of the power spectrum which then leads to higher concentration since the haloes are formed in the earlier epoch of the universe. In addition, the mass dependence of the halo concentrations for large mass haloes weaken as the the spectral index becomes more negative.

So far, we have discussed concentration-mass relations measured from N-body simulations and then calibrated using an analytical model. \citet{2016MNRAS.456.3068O} presented a theoretical concentration-mass relation using arguments that include the conservation of energy and the ellipsoidal collapse model of \citet{2001MNRAS.323....1S}. This relation agrees very well with simulation results in the realm where they exists and can be extrapolated to small mass regions, where we have no simulation data. The model shows a flattening of the halo concentrations at large masses. 

\begin{figure}[t]
\centering
\includegraphics[width=0.9\textwidth]{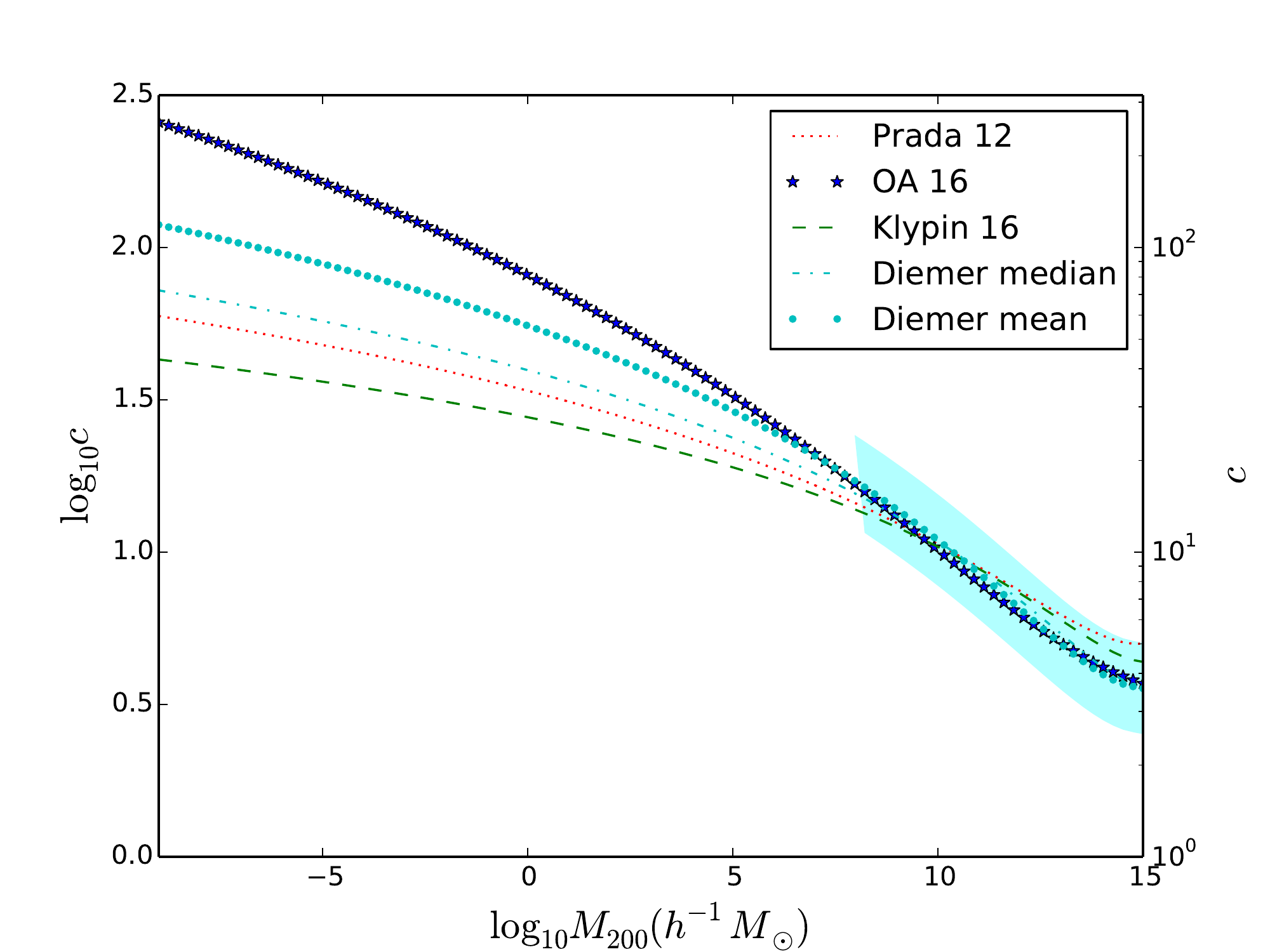}
\caption{The range of concentration for halos at  $z = 0$ for four different models of concentration-mass relations. The mass range of the cyan region shows the regions that have mostly been sampled by N-body simulations. The vertical range of the cyan region depicts the mean scatter in the simulated halo concentration.} 
\label{fig:conc_range}
\end{figure}

A comparison of a selected number of concentration-mass relations from the literature is shown in Figure \ref{fig:conc_range}. These is a great agreement between different relations in the simulated regions, but quite a disparity when these relations are extrapolated to low masses. 

\section{The evolution of the halo mass-concentration relation}
In this section, we will present the evolution of the halo concentration-mass relation with redshift. Assuming that the virial radius evolves as $R_{vir} \propto \frac{1}{1+z}$ (due to the scaling of the matter density with redshift for a fixed mass) ad that the scale radius remains constant, then the concentration is expected to $c \propto \frac{1}{1+z}$ \citep{2001MNRAS.321..559B}.  However, in recent years, simulations have confirmed that the evolution of the halo concentration is more complex than such a simple form. The mass-concentration relation becomes shallower with increasing redshift \citep{2016MNRAS.459.2106A}. This feature implies that at fixed mass, the concentration-mass relation evolves faster at lower redshifts \citep{2011MNRAS.411..584M,2015MNRAS.452.1217C} and is being attributed to the domination of dark energy at low redshifts. In fact, small masses evolve more strongly with redshift than large masses \citep{0004-637X-568-1-52,2003ApJ...597L...9Z,2011MNRAS.411..584M,2015MNRAS.452.1217C,2016MNRAS.456.3068O}.

Halo concentrations from high-redshift ($z=32$) simulations of microhaloes yield a nearly mass-independent concentration \citep{2014ApJ...788...27I}. The best fit form to the profile is the generalized NFW (gNFW) profile given by 
\begin{equation}
\rho_{\rm gNFW} = \frac{\rho_s}{\left(\frac{r}{r_s}\right)^{\alpha}\left(1 + \left(\frac{r}{r_s}\right)^{\beta}\right)^{\gamma}},
\end{equation}
where $\alpha$, $\beta$ and $\gamma$ are free parameters that are varied to fit the microhaloes. The measured halo concentrations were converted to those expected from an NFW profile for clear comparison. When extrapolated to $z=0,$ these concentrations are clearly lower than concentrations expected from power-law extrapolations of the mass-concentration relation. This result confirms that the power-law expressions of concentrations in terms of mass isn't the most physical and universal way to express halo concentrations. 

\section{Halo concentrations from the Einasto density profile}
Halo concentrations were traditionally defined using the NFW profile and have been studied extensively in the literature. \citet{2008MNRAS.390L..64D} first studied halo concentrations from the Einasto profile. The results show that fitting the same set of haloes to the NFW profile and the Einasto profile resulted in higher normalizations and steeper slope for the Einasto profile relative to the NFW profile. This is true for all definitions of concentrations, for the redshifts studied ($0 > z > 2$), and for both the full and relaxed haloes. In investigating the relation between the form of the MAH and the density profile, \citet{2013MNRAS.432.1103L} indicated that some MAHs were not good fits to the NFW profile. Fitting both the MAH and density profile to the Einasto profile resulted in better fits but for different values of $\alpha.$ The value of $\alpha$ closer to the NFW profile is 0.18. Profiles with $\alpha$'s higher than this for a fixed concentration seem to assemble faster than an NFW halo while profiles with $\alpha$'s lower than this assemble slower. Thus, in addition to characterizing an NFW halo by the concentration, the $\alpha$ parameter is also necessary to characterize an Einasto profile. Haloes with higher than average concentrations have lower values of $\alpha$ and vice versa for a given mass. Contrary to the upturn observed in \citet{2014arXiv1411.4001K} at the high mass end, \citet{2008MNRAS.390L..64D} find that the measured concentrations using the NFW profile or the Einasto profile are within $15\%$ for consistent Einasto shape parameters 

\section{Halo concentration and cosmological parameters}
Halo concentrations are dependent on the parameters of a the cosmological model. Cosmologies with lower values of $\sigma_8$ and $\Omega_m$ have been found to have lower concentrations relative to cosmologies with higher values of the same parameters \citep{2008MNRAS.391.1940M,2008MNRAS.390L..64D}. \citet{2008MNRAS.391.1940M} investigated this effect using simulations that were run with the parameters of the WMAP1, WMAP3 and WMAP5 cosmology. Cosmologies with lower $\Omega_m$ and $\sigma_8$ are expected to assemble haloes later for a given halo mass. Since the assembly time of a halo is related to the concentration, halo concentrations have different values dependent on the assumed cosmology. Halo concentrations increase with higher spectral indexes, $n$. 

In an effort for a more universal mass-concentration relation with less dependence on the assumed cosmology, halo concentrations may be expressed in terms of the rms (root-mean-square) of the matter density field ($\sigma(M)$) or the peak height parameter ($\nu$). Although the shape of the c-$\nu$ relation is constant with redshift, the amplitude has a redshift dependence that may be expressed as $D(z)^{0.5}$ \citep{2013ApJ...766...32B}.

\section{Halo concentrations and observational constraints. }
Due to the dependence of halo concentrations on cosmological parameters, attempts have ben made to constrain the cosmological parameters through observed shape and normalization of the mass-concentration relation. In particular, the dependence on the normalization of the matter power spectrum ($\sigma_8$) has been explored in the literature \citep{2000ApJ...529L...5W,2001ApJ...554..114E,2003MNRAS.340.1199H}. Relating the halo concentrations to the power spectrum of matter fluctuations, \citet{2001ApJ...554..114E} attempted to solve the over-concentration problem of $\Lambda CDM$ haloes via a change in the normalization of the power spectrum or the shape of the power spectrum or introducing a new form of dark matter particle. Similarly, \citet{2003MNRAS.340.1199H} related the halo concentrations to the full nonlinear matter power spectrum using the Halo Model of structure formation \citep{2002PhR...372....1C}, the halo mass function and a halo density profile. Since the nonlinear power spectrum varies with the mass-concentration relation, the halo mass-concentration relation may be constrained using the full nonlinear power spectrum in the regimes attainable with simulations and then extrapolated to predict the nonlinear power spectrum in the regimes outside the reach of simulations. The exercise wasn't really fruitful as the results weren't quite good beyond $k \sim 40 \rm{h} \rm{Mpc}.$ Alternatively, the observed concentrations may be minimized with a given theoretical model \citep{2010A&A...524A..68E}. In doing this, the degeneracy between $\sigma_8$ and $\Omega_m$ can be broken using the cosmic baryon fraction which then fixes the cosmological parameters. For all observational constraints of the mass-concentration relation, one should be wary of the effect of baryons on the measured observed concentrations (See section \ref{sec:obs_conc} for more details).   

Finally, the best fit cosmological parameters may be constrained from the shear peak abundance of weak lensing (peak counts) and a given normalization, slope and scatter of the concentration-mass relation \citep{2014JCAP...08..063M,2015A&A...574A.141C}. There is actually a degeneracy between the mass-concentration relation and the cosmological parameters ($\sigma_8$ and $\Omega_m$). Thus, one has to know one or the other precisely. 

\section{Scatter in halo concentrations}
\label{sec:scatter}
The distribution in halo concentrations have been found to follow a log-normal distribution, i.e a Gaussian distribution in the log of concentrations. First pointed out by \citet{0004-637X-535-1-30} and corroborated by a lot of other authors such as \citet{2007MNRAS.381.1450N,2008MNRAS.391.1940M}, the log-normal distribution describes both relaxed and unrelaxed haloes, albeit with a different mean and scatter. Specifically, the unrelaxed have higher dispersion relative and lower mean to the relaxed haloes. In terms of $c_{vir}$, \citet{2001MNRAS.321..559B}  find the scatter in halo concentrations to be $\Delta(\log c_{vir}) = 0.18$ for distinct haloes. However, the scatter in concentration is independent of the definition of the concentration, redshift, mass  \citep{2015ApJ...799..108D} and the underlying cosmology \citep{2008MNRAS.391.1940M}. The scatter in the log of concentration, $\sigma_{\ln c}$ for relaxed haloes ranges from  0.11 -- 0.27 \citep{0004-637X-535-1-30,0004-637X-568-1-52,2007MNRAS.381.1450N,2008MNRAS.390L..64D,2008MNRAS.391.1940M,2014MNRAS.441.3359D,2015ApJ...799..108D} irrespective of whether the fits are to the NFW density profile or Einasto profile.

Scatter in the concentrations for the full halo sample is $\sigma_{\ln c} = 0.25$ \citep{0004-637X-535-1-30} and $\sigma_{\ln c} = 0.30$ \citep{2008MNRAS.391.1940M}. However, \citet{2008MNRAS.390L..64D} find quite lower scatters in $\ln_c$ is given by $\sigma_{\ln c_{200}} = 0.15$ and $0.17$ for full sample of haloes fitted with the NFW profile and the Einasto profile respectively. Thus, the inclusion of the unrelaxed haloes skews the distribution to lower concentrations. 

Although the scatter in halo concentrations is generally found to be independent of mass \citep{2001MNRAS.321..559B,2008MNRAS.391.1940M,2015ApJ...799..108D}, some results indicate a weak mass dependence -- decreasing scatter with mass\citep{0004-637X-568-1-52,2007MNRAS.381.1450N,2008MNRAS.391.1940M,2016MNRAS.456.3068O}. This trend might indicate the uniformity of formation times for massive haloes and quite a disparity in collapse times for the less massive haloes.  

Sources of scatter in halo concentrations are thought to be from two sources -- the intrinsic spread in estimated halo concentrations from Poisson noise due to finite amount of haloes in a mass bin and the poor resolution of the halo structure, especially for small mass haloes \citep{2001MNRAS.321..559B}. Physical explanations for the scatter in the halo concentrations include the different collapse epochs of haloes that are in different environments. In order words,  the scatter is attributed to the variations in the MAH of the haloes. In particular, \citet{2007MNRAS.381.1450N} shows that the scatter is better accounted for if the the formation time of a halo is defined incorporating the history of all progenitors rather than the history of only the main progenitor. 

Although the scatter in the halo concentrations have been traditionally fit to the lognormal distributions, \citet{2013ApJ...766...32B} argue that the dispersion of concentrations can also be fit by a Gaussian distribution. This observation has been earlier pointed out by \citet{2008MNRAS.391.1940M}

From observations, the scatter in the halo concentrations has also been shown to be lognormal \citep{2012MNRAS.421.1073B,2016A&A...590A.126A} and decreasing with increasing mass. Additionally, the distribution in halo concentrations is an interesting way to get the fraction of galaxies/clusters with a given concentration to aid with a fair comparison to concentrations measured from observations. 

\section {Concentration and subhaloes}
\label{sec:subhaloes}
Not only at subhaloes of a given mass more concentrated than their distinct counterparts \citep{2001MNRAS.321..559B, 2005ApJ...634...51A}, their mass-concentration relations also have stronger mass dependences \citep{2001MNRAS.321..559B}. In a detailed study of the structure of subhaloes by \citet{2017MNRAS.466.4974M}, the concentration-mass relation of subhaloes was found to be quite dependent on the distance of the subhalo from the host halo centre. Subhaloes closer to the centre of the host halo have higher concentrations. Subhaloes also have higher dispersion/scatter relative to distinct haloes \citep{2005ApJ...634...51A} and are given as $\sigma_{\log_{10}c} =0.11 - 0.13$ \citep{2017MNRAS.466.4974M} and $\sigma_{\ln c_{vir}} = 0.24$ \citep{2001MNRAS.321..559B}. Physically, subhaloes are more concentrated than the distinct haloes due to the effects of interactions and tidal stripping as the subhalo orbits in the parent halo. The density structure is stripped in the outer parts while leaving the inner denser part of the subhalo.

\section {Effects of the environment on the concentration of haloes}
One important question to be answered in this section is what role does the environment -- such as a high density cluster environment or a low density void environment --  play on the concentration of haloes? Is it reasonable to expect higher concentrations for haloes in a region with high local density or vice versa? While the concentration of haloes may be independent of the local environment \citep{1999MNRAS.302..111L}, haloes in cluster environments may actually be more concentrated than haloes in the field for masses $\leq 5 \times 10^{11}h^{-1}M_{\odot}$ \citep{2005ApJ...634...51A}. However, parent haloes from cluster, field and void environments do not show any significant difference in their halo concentrations. This finding suggests that the difference in concentrations may be due to the presence of subhaloes. Recall that the subhaloes are usually more concentrated than haloes as discussed in section \ref{sec:subhaloes}. In addition, haloes in denser environments exhibit more scatter relative to haloes in less dense environments. Since the concentration of haloes is related to the formation time of the haloes, this points to the fact that the denser regions collapse and assemble their haloes earlier relative to the less dense regions. Therefore, halo formation time is linked to the global environment.  

Given that halo concentrations are related to the formation times of the haloes, formation times are linked to the global environment, then the concentrations should be affected by the global environment. This leads to the study of the dependence of the concentration on the halo assembly bias. The halo assembly bias is the dependence of halo clustering with formation time i.e the clustering of a haloes at a given mass is dependent on the assembly history of the haloes. Studying the correlation between concentration, clustering and halo formation time reveal that younger haloes are more clustered than their older counterparts for haloes that are ten times higher than the nonlinear scale, $M_*$ while the older haloes are more clustered relative to the younger haloes for haloes whose masses are lower than the nonlinear mass scale \citep{2007ApJ...657..664J}. Thus, the clustering bias of halos as a function of mass scales differently for both the older and the younger population. The younger population have a steeper slope while slope of the older population is shallower. The lines of the slope intersect around $M_*.$ Findings reveal that haloes with higher concentrations are more clustered than haloes with lower concentrations for masses less than the characteristic nonlinear mass scale, $M_*$ while haloes with higher concentrations were found to be less clustered for for haloes whose masses are greater than $M_*$. Given these findings, the statistics of haloes should be dependent on other halo properties other than the mass. The halo bias have been parametrized in terms of halo concentration and redshift for a fixed mass \citep{2006ApJ...652...71W}. This dependence is weak for haloes with masses $M>M_*,$ which makes the halo mass the dominant variable for estimating the halo bias.

Observationally, for the same stellar/halo mass, red galaxies are more concentrated than the blue galaxies \citep{2013MNRAS.428.2407W}. This observation corroborates the theoretical expectation between halo concentrations and halo assembly bias. 

\section{Observational halo concentration}
\label{sec:obs_conc}

There have been lots of concerted efforts to measure the concentrations of haloes from observations of dwarfs, galaxies or clusters. Measuring halo concentrations stem from a number of methods and assumptions such as using X-ray clusters through the assumption of hydrostatic equilibrium and spherical distribution; from weak lensing via the elongation and the tangential distortion of a background galaxy; from strong lensing via the strong distortion of the images of a background galaxy; and galactic dynamics that uses the measurement of the line of sight velocity as a function of radius. In this section, we will examine the results of halo concentrations measured from observations from the above-listed methods. It is important to note that halo concentrations from N-body simulations, until recently, are measured from the dark matter component only while observational concentrations are from mass measurements that do not separate the baryons from the dark matter.  To first approximation, assuming the mass of a galaxy/cluster is dominated by the dark matter component is fine.
\subsection{Halo concentrations from X-rays}
Measuring concentrations from X-ray temperature profiles involve the assumption of a spherical distribution and hydrostatic equilibrium. The equation for hydrostatic equilibrium is given by 
\begin{equation}
\frac{d (n(r) kT(r))}{dr}  = - \frac{G \mu m_p M(<r)n(r)}{r^2},
\end{equation}
where $G$, $\mu$, $m_p$, $n(r)$, $T(r)$, $M(<r)$ are the gravitational constant, mean molecular weight, mass of the proton, gas number density profile, gas temperature profile, and the enclosed total mass within a radius respectively.  Assuming the hydrostatic equilibrium, concentration estimates require the knowledge of a mass profile (NFW or Einasto in this case), a temperature profile and a gas density profile. The gas density profile may be estimated from the surface brightness profile. \citet{2007ApJ...664..123B} find consistent relations between observed concentrations and simulated halo concentrations using X-ray measurements for groups and clusters in the range $10^{12} \leq M \leq 10^{14} M_{\odot}.$ \citet{2016A&A...590A.126A} (\citet{2010A&A...524A..68E}), on the other hand, find that their observed halo concentrations have slopes that are higher (lower) relative to the theoretical expectation.

Limitations to the measurement of halo concentrations from X-ray profiles include the validity of hydrostatic equilibrium for galaxies and clusters. This assumption presumes that extra sources of pressure such as the magnetic field, turbulence, or cosmic rays are minimal relative to the thermal pressure of the electrons. This assumption was investigated by \citet{2003ASSL..281...87B}, who finds that the assumption holds on the cluster scales. There are also concerns regarding the assumed spherical distribution of particles. Clusters are mostly triaxial in shape; thus making them spherical along the major axis could lead to bias in the estimated concentrations. 

\subsection{Halo concentrations from weak and strong lensing measurements}
Using strong lensing to measure halo concentrations depends on the extreme distorted images of a background galaxy while weak lensing is based on the elongation and tangential distortion of a background galaxy. Lensing distortions are sensitive to the total gravitating mass of the cluster, which is dominated by dark matter, and is insensitive to the baryonic mass fraction. Concentrations may then be estimated from the shear or convergence map -- from weak lensing measurements or the Einstein radius -- for strong lensing estimates. Lensing measurements usually give estimates of halo properties in projection. These measurements are then deprojected to yield their 3D counterparts. Constraining the mass-concentration relation from observations may be effective using halo counts from weak lensing \citep{2011MNRAS.416.2539K} or the lensing convergence power spectrum \citep{2011MNRAS.416.2539K}. To make these constraints, assumptions are made for the form of the projected mass profile, the lensing shear or convergence profile and the cosmological parameters for the given cosmology e.g $\sigma_8$, $\Omega_m$. For a single galaxy, images from weak lensing are only fairly distorted; a significant signal becomes apparent when these distorted images are statistically averaged. Thus, concentration estimates using weak lensing analysis are based on stacking a number of galaxies. Stacking on weakly lensed clusters to give a statistical average of the dark matter profile have been found to quite effective and more accurate for comparing to theoretical estimates from simulations since profiles  from simulations are also averaged. This results in halo concentrations consistent with theoretical predictions \citep{2008JCAP...08..006M,2013MNRAS.434..878S,2013ApJ...769L..35O}. Another advantage to stacking clusters is that it corrects for uncorrelated large scale structure along the line of sight and the intrinsic ellipticity of the individual clusters \citep{2013ApJ...769L..35O}.

Even though some authors find consistent results with simulations as indicated above, others find that the normalization and slope of the  concentration-mass relation is biased low relative to the theoretical expectation \citep{2012MNRAS.421.1073B}. Concerns that may lead to such discrepancy include the triaxial shape and orientation of the clusters rather than assuming a symmetric spherical distribution, projection effects, cluster selection function, presence of substructure, the offset of the assumed centre, or the intrinsic ellipticity of the clusters. Other sources of bias in measuring of the halo concentrations include the presence of substructure (biases the concentration by about $5\%$) and the presence of a BCG in the centre of the cluster \citep{2012MNRAS.426.1558G}. Accounting for some of these biases leads to better agreement between the observed concentrations and the theoretical concentrations as demonstrated in \citet{2012MNRAS.419.3280S}, which accounts for the triaxial shape and orientation of the clusters; \citet{2015ApJ...806....4M,2014ApJ...797...34M}, who account for projection effects and the selection function of the clusters; \citet{2014ApJ...785...57D}, who verify that accounting for selection function leads to better estimates of the halo concentrations; the effect of the centre offset was found to be subdominant on the estimated concentrations while the shape noise (intrinsic ellipticity of the clusters) have a large effect on the estimated halo concentrations \citep{2014ApJ...785...57D}; \citet{2015ApJ...814..120D} found consistent slope with a higher normalization relative to the theoretical expectation when the impact of shape noise, centre offset and projection effects are accounted for in their weak lensing analysis; correcting for elongation of the cluster along the line of sight leads to unbiased mass estimates \citep{2012MNRAS.426.1558G} and slightly low concentration estimates. Even correcting for lensing bias does not always guarantee that observational concentrations will agree with theoretical expectations. This is evident in \citet{2016A&A...590A.126A} where weak lensing analysis was carried out on strong lensing selected clusters. 

The Einstein radius measured using strong lensing statistics may be used as a proxy for estimating the halo concentrations as in \citet{2011ApJ...737...74G}. Halo concentrations from strongly lensed clusters are usually higher than those from the X-ray clusters \citep{2000ApJ...529L...5W}. This is due to the fact that strongly lensed clusters are preferentially more concentrated. This effect is evident in the results of \citet{2007MNRAS.379..190C}, who find a higher slope and normalization for the concentration-mass relation. Similarly, To confirm that galaxies selected via their strong lensing signal are a biased sample, \citet{2014A&A...572A..19F} performed weak lensing analysis by stacking on a sample of strong lensing selected galaxies. The stacking process is expected to limit bias such as intrinsic ellipticity, substructures in the individual galaxies, and uncorrelated large scale structure. The resulting mass-concentration relation was found to be higher in normalization and steeper in slope relative to expectations from theoretical concentrations. Even though \citet{2007ApJ...654..714H} showed that strongly lensed clusters are $18\%$ more concentrated relative to the overall population and that the 2D concentration is $34\%$ more concentrated relative to the total population, correcting for this bias doesn't seem to solve the over-concentration problem of observed clusters 

Concerns regarding using strong lensing for measuring halo concentrations include the fact that strong lensing analysis are usually limited to the inner part of the cluster and do not extend to the outer parts of the halo; and the orientation of the clusters. Indeed, the strong lensing analysis of this sample of X-ray selected clusters showed that the clusters were preferentially aligned along the line of sight \citep{2012MNRAS.419.3280S}. Combining both strong and weak lensing analysis of clusters yield better estimates of both the strongly-lensed central region and the weakly-lensed outer region of a cluster. This leads to an excellent reconstruction of the mass profile. 

\subsection{Halo concentrations from galactic dynamics}
Measuring the concentration of haloes from the orbits of galaxies in clusters have been used as a method of galactic dynamics for estimating the mass and concentrations of clusters. Variants of galactic dynamics include the Jean's analysis, the virial theorem, the projected phase-space analysis using an anisotropic model of the distribution function. Estimates from galactic dyanamics assumes a spherical distribution of particles, the Jeans equation and the line of sight velocity profile. As investigated by \citet{2010MNRAS.408.2442W} using a model for the distribution function, the projected phase space analysis yields higher concentrations relative to the theoretical concentrations, for realistic fixed cosmological parameters.  


\section{Discussion on theoretical and observed results}
Comparison between theoretical halo concentrations measured from simulations and observational halo concentrations show agreement \citep{2008JCAP...08..006M, 2013MNRAS.434..878S, 2013ApJ...769L..35O}. A comparison to N-body simulation results and results from observations -- such as weak and strong lensing, X-rays and galaxy dynamics -- have been reported in \citet{2013ApJ...766...32B} and were found to be in good agreement with each other.  In some cases, however, there are disagreements (e.g in \citet{2010A&A...524A..68E}) as discussed above. Given the various means of estimating halo concentrations from observations, we will discuss below the assumptions and concerns related to these measurements that could lead to discrepancies. These assumptions include hydrostatic equilibrium in estimates made from X-ray clusters and the dynamical state of clusters in virial equilibrium in the estimates made from galactic dynamics. Although measurements from gravitational lensing are not plagued by a priori assumptions, they are majorly affected by projection effects. 

Major concerns related to the measurement of the concentration-mass relation include the intrinsic ellipticity of galaxies, the non-spherical distribution of matter, projection effects, the presence of baryons, the orientation of the individual clusters, the selection of the sample of galaxies/clusters, the method of reconstructing the masses, the presence of large scale structure along the line of sight, the number of samples, the radial range for fitting the concentration parameters, and the mass range for estimating the mass-concentration relation. The methods are affected by some of these concerns in different ways. In no particular order, these concerns are briefly discussed below:
\begin{itemize}

\item{Hydrostatic equilibrium --  As previously discussed, hydrostatic equilibrium is usually assumed in the estimation cluster properties from X-ray clusters. There are concerns that these clusters may not actually be in equilibrium and possibility of other sources of pressure such as magnetic pressure, turbulence and cosmic ray pressure, in addition to the thermal pressure. However, this assumption has been found to be compatible with galaxies and groups when compared to stellar dynamics \citep{2006MNRAS.373..157B, 2006ApJ...646..899H} and clusters when compared to lensing \citep{1996ApJ...469..494E, 1999ApJ...520L..21M, 2003ASSL..281...87B}}

\item{Orientation and non-spherical matter distribution -- The shape of haloes, and also clusters, are mostly triaxial, which makes the assumption of sphericity a not-so-good approximation. Assumptions of sphericity leads to an over-prediction or under-prediction of the halo concentrations depending on the orientation of the cluster along the line of sight. Elongation along the line of sight over-predicts the halo concentrations while elongation along the plane of the sky under-predicts halo concentrations.Thus, the triaxial mass distribution has to be incorporated into the analysis of clusters. Assuming an NFW-ellipsoid halo with seven parameters -- the mass, concentration, three orientation angles and two axial ratios, halo concentrations were estimated from observations tend to be lower concentrations ($c \leq 3$) and are consistent with concentrations from theoretical expectations \citep{2012MNRAS.419.3280S}}.

\item{Baryons  -- Traditionally, theoretical measurements of concentrations are made from N-body simulations of dark matter particles. However, galaxies and clusters in observations have contributions from baryons -- stars, hot gas, and cold gas -- and dark matter. Clusters, in particular, are dominated by hot gas while the groups have more stars than gas. The stellar and cold gas fraction increases to a maximum mass around $M_* = 10^{13}h^{-1}M_{\odot}$ before it starts decreasing. The hot gas component, on the other hand, keeps increasing with increasing mass. These baryons, however, are expected to be dominated in the innermost region of a galaxy/cluster. Resolution effects in simulations thwart the inclusion of such inner most region in simulations. Nevertheless, the presence of the baryons is expected to modify either the dark matter distribution or its properties such as the concentration. 

Baryonic effects -- such as gas cooling and feedback from supernovas and AGNs  --  only lead to about $10\%$ difference in the halo concentrations when the innermost $10\%$ region of the halo isn't considered \citep{2010MNRAS.405.2161D}. On the other hand, including radiative effects such as cooling, star formation, and feedback in simulations lead to about $20-50\%$ larger concentrations \citep{2013ApJ...776...39R} relative to dark matter only simulations. \citet{2012MNRAS.424.1244F} made efforts to review the effects of baryon cooling on observed concentrations when compared to theoretical concentrations. The author adopts a semi-analytic model to investigate the effects of baryonic physics on the concentration-mass relation of groups and clusters. In particular, the effect of adiabatic contraction -- the response of dark matter to the presence of baryons -- and other baryonic and gas physics on the observational halo concentrations were investigated. To this end, an initial dark matter profile is expected to disrupted by adiabatic cooling. This involves the cooling of the gas and the formation of stars which drags the dark matter closer to the centre making the halo more concentrated. It is important to note, however, that the role played by adiabatic contraction is more effective on the small mass systems than massive systems. For realistic stellar mass fractions, the effect of adiabatic contraction on the concentration-mass relation is minimal. }

\item{Mass reconstruction -- Mass estimates of observed galaxies/clusters are usually based on proxies, which include the temperature, Sunyaev-Zeldovich (SZ) Compton Y parameter \citep{2011ApJ...737...74G}, velocity dispersion \citep{2010MNRAS.408.2442W}, X-ray luminosity or using lensing analysis. In particular, mass reconstruction via strong lensing may not be able to accurately determine the total cluster mass effectively due to the fact that it samples only the innermost part of a cluster. Calibrations between these proxies and masses are subject to scatter, which is usually not accounted for in the mass reconstruction. The masses from these proxies suffer from different types of bias. For example, X-ray masses are usually biased low relative to weak lensing masses by about $10\%$ or $25\% $ \citep{2013ApJ...767..116M} or \citep{2012NJPh...14e5018R} which leads to higher concentrations. 

Still on the reconstruction of the galaxy/cluster masses, the assumption of a given mass profile may plague the mass estimated. As previously discussed, it is expected that the original mass profile is evolves to a new profile due to adiabatic contraction. Commonly, the NFW profile is assumed. Probably, the effects of baryons modifies the density distribution to a different distribution.}

\item{Sample selection --  Samples are usually selected based on some intrinsic property. These include the strong lensing signal, X-ray morphology, X-ray luminosity etc. This leads to a from of bias in the selected sample. Sample selection based on the strong lensing signal tend to lead biased halo properties since it tends to select the most concentrated haloes; Sample selection based on X-ray morphology lead to less bias \citep{2014ApJ...797...34M} relative to selection based on X-ray luminosity, which leads to higher normalization of the concentration-mass relation. These selection effects tend to suggest that observations skewed towards the high end of the concentration distribution. Given that the scatter in concentrations at a given mass is quite big, the samples selected from observations may actually be from the higher end of the distribution. Reconciling observed concentrations will also involve taking into account all sources of bias into consideration.}

\item{Radial range -- \citet{2013ApJ...766...32B} find that estimated halo concentrations depend on the range of radial fitting for the density profile. Including the innermost regions of the halo tends to lead to lower concentrations by about $5\%.$ Other radial variations make only a maximum difference of about $10\%$ in the measured concentrations. On the other hand, the radial range for fitting the halo concentrations investigated by \citet{2013ApJ...776...39R} shows that reducing the radial range to include the innermost parts of the cluster leads to concentration-mass relations with a higher slope and higher normalization. It is important to note that these changes are minimal.}

\item{Projection effects -- Weak lensing analysis of clusters measure the cluster properties in projection. Thus, there are concerns regarding the de-projection of the 2D measured properties into 3D properties \citep{2008JCAP...08..006M}. This de-projection usually assumes a spherical distribution; however, galaxies/clusters are known to be aspherical in shape. Depending on the elongation of the main axis (along the line of sight or along the plane of the sky), estimated 3D concentrations may be overestimated or underestimated respectively.}

\item{Others -- Other effects that could contribute to observational systematics include: sample size -- the observed concentrations are based on a sample of clusters of the order of ten \citep{2013ApJ...766...32B} whereas simulations typically have larger samples; mass range -- the range of masses for these samples are quite limited and thus are not suitable for constraining the concentration-mass relation; intrinsic ellipticity of galaxies/clusters -- the triaxial nature of galaxies/clusters creates an intrinsic shape property that should properly be accounted for in weak lensing tangential shear/convergence and strong lensing arcs. }

\end{itemize}

In conclusion, concentration estimates from observations are generally marred with systematics that need to be properly accounted when comparing to theoretical results. A good review of observational halo concentrations measured from different methods is found in \citet{2016MNRAS.455..892G}. \citet {2007ApJ...664..123B} also touches on a number comparisons regarding observed halo concentrations. As with the theoretical concentrations from simulations, the scatter in observed halo concentrations at a fixed mass also follow a log-normal distribution \citep{2007ApJ...664..123B}. Although we made great efforts to review relevant literature to the topic of interest, we do not claim to have reviewed all articles in this field and apologize in advance for any omission. 

\section*{Acknowledgement}
The author thanks James E. Taylor as this review emanated from a discussion with him. This work is supported by the University of Waterloo and the Perimeter Institute. Research at the Perimeter Institute is supported in part by the Government of Canada through Innovation, Science and Economic Development Canada and by the Province of Ontario through the Ministry of Research, Innovation and Science. 

\section*{References}

\bibliographystyle{elsarticle-harv}

\bibliography{conc_review_final.bib}

\begin{thebibliography}{76}
\expandafter\ifx\csname natexlab\endcsname\relax\def\natexlab#1{#1}\fi
\expandafter\ifx\csname url\endcsname\relax
  \def\url#1{\texttt{#1}}\fi
\expandafter\ifx\csname urlprefix\endcsname\relax\def\urlprefix{URL }\fi

\bibitem[{{Amodeo} et~al.(2016){Amodeo}, {Ettori}, {Capasso}, and
  {Sereno}}]{2016A&A...590A.126A}
{Amodeo}, S., {Ettori}, S., {Capasso}, R., {Sereno}, M., May 2016. {The
  relation between mass and concentration in X-ray galaxy clusters at high
  redshift}. aap 590, A126.

\bibitem[{{Angel} et~al.(2016){Angel}, {Poole}, {Ludlow}, {Duffy}, {Geil},
  {Mutch}, {Mesinger}, and {Wyithe}}]{2016MNRAS.459.2106A}
{Angel}, P.~W., {Poole}, G.~B., {Ludlow}, A.~D., {Duffy}, A.~R., {Geil}, P.~M.,
  {Mutch}, S.~J., {Mesinger}, A., {Wyithe}, J.~S.~B., Jun. 2016. {Dark-ages
  reionization and galaxy formation simulation - II. Spin and concentration
  parameters for dark matter haloes during the epoch of reionization}. mnras
  459, 2106--2117.

\bibitem[{{Avila-Reese} et~al.(2005){Avila-Reese}, {Col{\'{\i}}n},
  {Gottl{\"o}ber}, {Firmani}, and {Maulbetsch}}]{2005ApJ...634...51A}
{Avila-Reese}, V., {Col{\'{\i}}n}, P., {Gottl{\"o}ber}, S., {Firmani}, C.,
  {Maulbetsch}, C., Nov. 2005. {The Dependence on Environment of Cold Dark
  Matter Halo Properties}. apj 634, 51--69.

\bibitem[{{Bah{\'e}} et~al.(2012){Bah{\'e}}, {McCarthy}, and
  {King}}]{2012MNRAS.421.1073B}
{Bah{\'e}}, Y.~M., {McCarthy}, I.~G., {King}, L.~J., Apr. 2012. {Mock weak
  lensing analysis of simulated galaxy clusters: bias and scatter in mass and
  concentration}. mnras 421, 1073--1088.

\bibitem[{{Bhattacharya} et~al.(2013){Bhattacharya}, {Habib}, {Heitmann}, and
  {Vikhlinin}}]{2013ApJ...766...32B}
{Bhattacharya}, S., {Habib}, S., {Heitmann}, K., {Vikhlinin}, A., Mar. 2013.
  {Dark Matter Halo Profiles of Massive Clusters: Theory versus Observations}.
  apj 766, 32.

\bibitem[{{Bridges} et~al.(2006){Bridges}, {Gebhardt}, {Sharples}, {Faifer},
  {Forte}, {Beasley}, {Zepf}, {Forbes}, {Hanes}, and
  {Pierce}}]{2006MNRAS.373..157B}
{Bridges}, T., {Gebhardt}, K., {Sharples}, R., {Faifer}, F.~R., {Forte}, J.~C.,
  {Beasley}, M.~A., {Zepf}, S.~E., {Forbes}, D.~A., {Hanes}, D.~A., {Pierce},
  M., Nov. 2006. {The globular cluster kinematics and galaxy dark matter
  content of NGC 4649 (M60)}. mnras 373, 157--166.

\bibitem[{{Bullock} et~al.(2001){Bullock}, {Kolatt}, {Sigad}, {Somerville},
  {Kravtsov}, {Klypin}, {Primack}, and {Dekel}}]{2001MNRAS.321..559B}
{Bullock}, J.~S., {Kolatt}, T.~S., {Sigad}, Y., {Somerville}, R.~S.,
  {Kravtsov}, A.~V., {Klypin}, A.~A., {Primack}, J.~R., {Dekel}, A., Mar. 2001.
  {Profiles of dark haloes: evolution, scatter and environment}. mnras 321,
  559--575.

\bibitem[{{Buote}(2003)}]{2003ASSL..281...87B}
{Buote}, D.~A., Apr. 2003. {A Chandra and XMM View of the Mass {amp} Metals in
  Galaxy Groups and Clusters}. In: {Rosenberg}, J.~L., {Putman}, M.~E. (Eds.),
  The IGM/Galaxy Connection. The Distribution of Baryons at z=0. Vol. 281 of
  Astrophysics and Space Science Library. p.~87.

\bibitem[{{Buote} et~al.(2007){Buote}, {Gastaldello}, {Humphrey}, {Zappacosta},
  {Bullock}, {Brighenti}, and {Mathews}}]{2007ApJ...664..123B}
{Buote}, D.~A., {Gastaldello}, F., {Humphrey}, P.~J., {Zappacosta}, L.,
  {Bullock}, J.~S., {Brighenti}, F., {Mathews}, W.~G., Jul. 2007. {The X-Ray
  Concentration-Virial Mass Relation}. apj 664, 123--134.

\bibitem[{{Cardone} et~al.(2015){Cardone}, {Camera}, {Sereno}, {Covone},
  {Maoli}, and {Scaramella}}]{2015A&A...574A.141C}
{Cardone}, V.~F., {Camera}, S., {Sereno}, M., {Covone}, G., {Maoli}, R.,
  {Scaramella}, R., Feb. 2015. {Mass-concentration relation and weak lensing
  peak counts}. aap 574, A141.

\bibitem[{{Comerford} and {Natarajan}(2007)}]{2007MNRAS.379..190C}
{Comerford}, J.~M., {Natarajan}, P., Jul. 2007. {The observed
  concentration-mass relation for galaxy clusters}. mnras 379, 190--200.

\bibitem[{{Cooray} and {Sheth}(2002)}]{2002PhR...372....1C}
{Cooray}, A., {Sheth}, R., Dec. 2002. {Halo models of large scale structure}.
  physrep 372, 1--129.

\bibitem[{{Correa} et~al.(2015){Correa}, {Wyithe}, {Schaye}, and
  {Duffy}}]{2015MNRAS.452.1217C}
{Correa}, C.~A., {Wyithe}, J.~S.~B., {Schaye}, J., {Duffy}, A.~R., Sep. 2015.
  {The accretion history of dark matter haloes - III. A physical model for the
  concentration-mass relation}. mnras 452, 1217--1232.

\bibitem[{{Diemer} and {Kravtsov}(2015)}]{2015ApJ...799..108D}
{Diemer}, B., {Kravtsov}, A.~V., Jan. 2015. {A Universal Model for Halo
  Concentrations}. apj 799, 108.

\bibitem[{{Du} and {Fan}(2014)}]{2014ApJ...785...57D}
{Du}, W., {Fan}, Z., Apr. 2014. {Effects of Center Offset and Noise on
  Weak-Lensing Derived Concentration-Mass Relation of Dark Matter Halos}. apj
  785, 57.

\bibitem[{{Du} et~al.(2015){Du}, {Fan}, {Shan}, {Zhao}, {Covone}, {Fu}, and
  {Kneib}}]{2015ApJ...814..120D}
{Du}, W., {Fan}, Z., {Shan}, H., {Zhao}, G.-B., {Covone}, G., {Fu}, L.,
  {Kneib}, J.-P., Dec. 2015. {Mass-Concentration Relation of Clusters of
  Galaxies from CFHTLenS}. apj 814, 120.

\bibitem[{{Duffy} et~al.(2008){Duffy}, {Schaye}, {Kay}, and {Dalla
  Vecchia}}]{2008MNRAS.390L..64D}
{Duffy}, A.~R., {Schaye}, J., {Kay}, S.~T., {Dalla Vecchia}, C., Oct. 2008.
  {Dark matter halo concentrations in the Wilkinson Microwave Anisotropy Probe
  year 5 cosmology}. mnras 390, L64--L68.

\bibitem[{{Duffy} et~al.(2010){Duffy}, {Schaye}, {Kay}, {Dalla Vecchia},
  {Battye}, and {Booth}}]{2010MNRAS.405.2161D}
{Duffy}, A.~R., {Schaye}, J., {Kay}, S.~T., {Dalla Vecchia}, C., {Battye},
  R.~A., {Booth}, C.~M., Jul. 2010. {Impact of baryon physics on dark matter
  structures: a detailed simulation study of halo density profiles}. mnras 405,
  2161--2178.

\bibitem[{{Dutton} and {Macci{\`o}}(2014)}]{2014MNRAS.441.3359D}
{Dutton}, A.~A., {Macci{\`o}}, A.~V., Jul. 2014. {Cold dark matter haloes in
  the Planck era: evolution of structural parameters for Einasto and NFW
  profiles}. mnras 441, 3359--3374.

\bibitem[{{Eke} et~al.(2001){Eke}, {Navarro}, and
  {Steinmetz}}]{2001ApJ...554..114E}
{Eke}, V.~R., {Navarro}, J.~F., {Steinmetz}, M., Jun. 2001. {The Power Spectrum
  Dependence of Dark Matter Halo Concentrations}. apj 554, 114--125.

\bibitem[{{Ettori} et~al.(2010){Ettori}, {Gastaldello}, {Leccardi}, {Molendi},
  {Rossetti}, {Buote}, and {Meneghetti}}]{2010A&A...524A..68E}
{Ettori}, S., {Gastaldello}, F., {Leccardi}, A., {Molendi}, S., {Rossetti}, M.,
  {Buote}, D., {Meneghetti}, M., Dec. 2010. {Mass profiles and c-M$_{DM}$
  relation in X-ray luminous galaxy clusters}. aap 524, A68.

\bibitem[{{Evrard} et~al.(1996){Evrard}, {Metzler}, and
  {Navarro}}]{1996ApJ...469..494E}
{Evrard}, A.~E., {Metzler}, C.~A., {Navarro}, J.~F., Oct. 1996. {Mass Estimates
  of X-Ray Clusters}. apj 469, 494.

\bibitem[{{Fedeli}(2012)}]{2012MNRAS.424.1244F}
{Fedeli}, C., Aug. 2012. {The effects of baryonic cooling on the
  concentration-mass relation}. mnras 424, 1244--1260.

\bibitem[{{Fo{\"e}x} et~al.(2014){Fo{\"e}x}, {Motta}, {Jullo}, {Limousin}, and
  {Verdugo}}]{2014A&A...572A..19F}
{Fo{\"e}x}, G., {Motta}, V., {Jullo}, E., {Limousin}, M., {Verdugo}, T., Dec.
  2014. {SARCS strong-lensing galaxy groups. II. Mass-concentration relation
  and strong-lensing bias}. aap 572, A19.

\bibitem[{{Gao} et~al.(2008){Gao}, {Navarro}, {Cole}, {Frenk}, {White},
  {Springel}, {Jenkins}, and {Neto}}]{2008MNRAS.387..536G}
{Gao}, L., {Navarro}, J.~F., {Cole}, S., {Frenk}, C.~S., {White}, S.~D.~M.,
  {Springel}, V., {Jenkins}, A., {Neto}, A.~F., Jun. 2008. {The redshift
  dependence of the structure of massive {$\Lambda$} cold dark matter haloes}.
  mnras 387, 536--544.

\bibitem[{{Giocoli} et~al.(2012{\natexlab{a}}){Giocoli}, {Meneghetti},
  {Ettori}, and {Moscardini}}]{2012MNRAS.426.1558G}
{Giocoli}, C., {Meneghetti}, M., {Ettori}, S., {Moscardini}, L., Oct.
  2012{\natexlab{a}}. {Cosmology in two dimensions: the concentration-mass
  relation for galaxy clusters}. mnras 426, 1558--1573.

\bibitem[{{Giocoli} et~al.(2012{\natexlab{b}}){Giocoli}, {Tormen}, and
  {Sheth}}]{2012MNRAS.422..185G}
{Giocoli}, C., {Tormen}, G., {Sheth}, R.~K., May 2012{\natexlab{b}}. {Formation
  times, mass growth histories and concentrations of dark matter haloes}. mnras
  422, 185--198.

\bibitem[{{Gralla} et~al.(2011){Gralla}, {Sharon}, {Gladders}, {Marrone},
  {Barrientos}, {Bayliss}, {Bonamente}, {Bulbul}, {Carlstrom}, {Culverhouse},
  {Gilbank}, {Greer}, {Hasler}, {Hawkins}, {Hennessy}, {Joy}, {Koester},
  {Lamb}, {Leitch}, {Miller}, {Mroczkowski}, {Muchovej}, {Oguri}, {Plagge},
  {Pryke}, and {Woody}}]{2011ApJ...737...74G}
{Gralla}, M.~B., {Sharon}, K., {Gladders}, M.~D., {Marrone}, D.~P.,
  {Barrientos}, L.~F., {Bayliss}, M., {Bonamente}, M., {Bulbul}, E.,
  {Carlstrom}, J.~E., {Culverhouse}, T., {Gilbank}, D.~G., {Greer}, C.,
  {Hasler}, N., {Hawkins}, D., {Hennessy}, R., {Joy}, M., {Koester}, B.,
  {Lamb}, J., {Leitch}, E., {Miller}, A., {Mroczkowski}, T., {Muchovej}, S.,
  {Oguri}, M., {Plagge}, T., {Pryke}, C., {Woody}, D., Aug. 2011.
  {Sunyaev-Zel'dovich Effect Observations of Strong Lensing Galaxy Clusters:
  Probing the Overconcentration Problem}. apj 737, 74.

\bibitem[{{Groener} et~al.(2016){Groener}, {Goldberg}, and
  {Sereno}}]{2016MNRAS.455..892G}
{Groener}, A.~M., {Goldberg}, D.~M., {Sereno}, M., Jan. 2016. {The galaxy
  cluster concentration-mass scaling relation}. mnras 455, 892--919.

\bibitem[{{Hahn} and {Abel}(2011)}]{2011MNRAS.415.2101H}
{Hahn}, O., {Abel}, T., Aug. 2011. {Multi-scale initial conditions for
  cosmological simulations}. mnras 415, 2101--2121.

\bibitem[{{Hennawi} et~al.(2007){Hennawi}, {Dalal}, {Bode}, and
  {Ostriker}}]{2007ApJ...654..714H}
{Hennawi}, J.~F., {Dalal}, N., {Bode}, P., {Ostriker}, J.~P., Jan. 2007.
  {Characterizing the Cluster Lens Population}. apj 654, 714--730.

\bibitem[{{Hoffman} and {Shaham}(1985)}]{1985ApJ...297...16H}
{Hoffman}, Y., {Shaham}, J., Oct. 1985. {Local density maxima - Progenitors of
  structure}. apj 297, 16--22.

\bibitem[{{Huffenberger} and {Seljak}(2003)}]{2003MNRAS.340.1199H}
{Huffenberger}, K.~M., {Seljak}, U., Apr. 2003. {Halo concentration and the
  dark matter power spectrum}. mnras 340, 1199--1204.

\bibitem[{{Humphrey} et~al.(2006){Humphrey}, {Buote}, {Gastaldello},
  {Zappacosta}, {Bullock}, {Brighenti}, and {Mathews}}]{2006ApJ...646..899H}
{Humphrey}, P.~J., {Buote}, D.~A., {Gastaldello}, F., {Zappacosta}, L.,
  {Bullock}, J.~S., {Brighenti}, F., {Mathews}, W.~G., Aug. 2006. {A Chandra
  View of Dark Matter in Early-Type Galaxies}. apj 646, 899--918.

\bibitem[{{Ishiyama}(2014)}]{2014ApJ...788...27I}
{Ishiyama}, T., Jun. 2014. {Hierarchical Formation of Dark Matter Halos and the
  Free Streaming Scale}. apj 788, 27.

\bibitem[{Jing(2000)}]{0004-637X-535-1-30}
Jing, Y.~P., 2000. The density profile of equilibrium and nonequilibrium dark
  matter halos. The Astrophysical Journal 535~(1), 30.
\newline\urlprefix\url{http://stacks.iop.org/0004-637X/535/i=1/a=30}

\bibitem[{{Jing} et~al.(2007){Jing}, {Suto}, and {Mo}}]{2007ApJ...657..664J}
{Jing}, Y.~P., {Suto}, Y., {Mo}, H.~J., Mar. 2007. {The Dependence of Dark Halo
  Clustering on Formation Epoch and Concentration Parameter}. apj 657,
  664--668.

\bibitem[{{King} and {Mead}(2011)}]{2011MNRAS.416.2539K}
{King}, L.~J., {Mead}, J.~M.~G., Oct. 2011. {The mass-concentration
  relationship of virialized haloes and its impact on cosmological
  observables}. mnras 416, 2539--2549.

\bibitem[{{Klypin} et~al.(2014){Klypin}, {Yepes}, {Gottlober}, {Prada}, and
  {Hess}}]{2014arXiv1411.4001K}
{Klypin}, A., {Yepes}, G., {Gottlober}, S., {Prada}, F., {Hess}, S., Nov. 2014.
  {MultiDark simulations: the story of dark matter halo concentrations and
  density profiles}. ArXiv e-prints.

\bibitem[{{Klypin} et~al.(2011){Klypin}, {Trujillo-Gomez}, and
  {Primack}}]{2011ApJ...740..102K}
{Klypin}, A.~A., {Trujillo-Gomez}, S., {Primack}, J., Oct. 2011. {Dark Matter
  Halos in the Standard Cosmological Model: Results from the Bolshoi
  Simulation}. apj 740, 102.

\bibitem[{{Lemson} and {Kauffmann}(1999)}]{1999MNRAS.302..111L}
{Lemson}, G., {Kauffmann}, G., Jan. 1999. {Environmental influences on dark
  matter haloes and consequences for the galaxies within them}. mnras 302,
  111--117.

\bibitem[{{Ludlow} et~al.(2014){Ludlow}, {Navarro}, {Angulo}, {Boylan-Kolchin},
  {Springel}, {Frenk}, and {White}}]{2014MNRAS.441..378L}
{Ludlow}, A.~D., {Navarro}, J.~F., {Angulo}, R.~E., {Boylan-Kolchin}, M.,
  {Springel}, V., {Frenk}, C., {White}, S.~D.~M., Jun. 2014. {The
  mass-concentration-redshift relation of cold dark matter haloes}. mnras 441,
  378--388.

\bibitem[{{Ludlow} et~al.(2013){Ludlow}, {Navarro}, {Boylan-Kolchin}, {Bett},
  {Angulo}, {Li}, {White}, {Frenk}, and {Springel}}]{2013MNRAS.432.1103L}
{Ludlow}, A.~D., {Navarro}, J.~F., {Boylan-Kolchin}, M., {Bett}, P.~E.,
  {Angulo}, R.~E., {Li}, M., {White}, S.~D.~M., {Frenk}, C., {Springel}, V.,
  Jun. 2013. {The mass profile and accretion history of cold dark matter
  haloes}. mnras 432, 1103--1113.

\bibitem[{{Ludlow} et~al.(2012){Ludlow}, {Navarro}, {Li}, {Angulo},
  {Boylan-Kolchin}, and {Bett}}]{2012MNRAS.427.1322L}
{Ludlow}, A.~D., {Navarro}, J.~F., {Li}, M., {Angulo}, R.~E., {Boylan-Kolchin},
  M., {Bett}, P.~E., Dec. 2012. {The dynamical state and mass-concentration
  relation of galaxy clusters}. mnras 427, 1322--1328.

\bibitem[{{Macci{\`o}} et~al.(2008){Macci{\`o}}, {Dutton}, and {van den
  Bosch}}]{2008MNRAS.391.1940M}
{Macci{\`o}}, A.~V., {Dutton}, A.~A., {van den Bosch}, F.~C., Dec. 2008.
  {Concentration, spin and shape of dark matter haloes as a function of the
  cosmological model: WMAP1, WMAP3 and WMAP5 results}. mnras 391, 1940--1954.

\bibitem[{{Macci{\`o}} et~al.(2007){Macci{\`o}}, {Dutton}, {van den Bosch},
  {Moore}, {Potter}, and {Stadel}}]{2007MNRAS.378...55M}
{Macci{\`o}}, A.~V., {Dutton}, A.~A., {van den Bosch}, F.~C., {Moore}, B.,
  {Potter}, D., {Stadel}, J., Jun. 2007. {Concentration, spin and shape of dark
  matter haloes: scatter and the dependence on mass and environment}. mnras
  378, 55--71.

\bibitem[{{Mahdavi} et~al.(2013){Mahdavi}, {Hoekstra}, {Babul}, {Bildfell},
  {Jeltema}, and {Henry}}]{2013ApJ...767..116M}
{Mahdavi}, A., {Hoekstra}, H., {Babul}, A., {Bildfell}, C., {Jeltema}, T.,
  {Henry}, J.~P., Apr. 2013. {Joint Analysis of Cluster Observations. II.
  Chandra/XMM-Newton X-Ray and Weak Lensing Scaling Relations for a Sample of
  50 Rich Clusters of Galaxies}. apj 767, 116.

\bibitem[{{Mainini} and {Romano}(2014)}]{2014JCAP...08..063M}
{Mainini}, R., {Romano}, A., Aug. 2014. {Constraining the mass-concentration
  relation through weak lensing peak function}. jcap 8, 063.

\bibitem[{{Mandelbaum} et~al.(2008){Mandelbaum}, {Seljak}, and
  {Hirata}}]{2008JCAP...08..006M}
{Mandelbaum}, R., {Seljak}, U., {Hirata}, C.~M., Aug. 2008. {A halo
  mass-concentration relation from weak lensing}. jcap 8, 006.

\bibitem[{{Mathiesen} et~al.(1999){Mathiesen}, {Evrard}, and
  {Mohr}}]{1999ApJ...520L..21M}
{Mathiesen}, B., {Evrard}, A.~E., {Mohr}, J.~J., Jul. 1999. {The Effects of
  Clumping and Substructure on Intracluster Medium Mass Measurements}. apjl
  520, L21--L24.

\bibitem[{{Meneghetti} and {Rasia}(2013)}]{2013arXiv1303.6158M}
{Meneghetti}, M., {Rasia}, E., Mar. 2013. {Reconciling extremely different
  concentration-mass relations}. ArXiv e-prints.

\bibitem[{{Meneghetti} et~al.(2014){Meneghetti}, {Rasia}, {Vega}, {Merten},
  {Postman}, {Yepes}, {Sembolini}, {Donahue}, {Ettori}, {Umetsu}, {Balestra},
  {Bartelmann}, {Ben{\'{\i}}tez}, {Biviano}, {Bouwens}, {Bradley},
  {Broadhurst}, {Coe}, {Czakon}, {De Petris}, {Ford}, {Giocoli},
  {Gottl{\"o}ber}, {Grillo}, {Infante}, {Jouvel}, {Kelson}, {Koekemoer},
  {Lahav}, {Lemze}, {Medezinski}, {Melchior}, {Mercurio}, {Molino},
  {Moscardini}, {Monna}, {Moustakas}, {Moustakas}, {Nonino}, {Rhodes},
  {Rosati}, {Sayers}, {Seitz}, {Zheng}, and {Zitrin}}]{2014ApJ...797...34M}
{Meneghetti}, M., {Rasia}, E., {Vega}, J., {Merten}, J., {Postman}, M.,
  {Yepes}, G., {Sembolini}, F., {Donahue}, M., {Ettori}, S., {Umetsu}, K.,
  {Balestra}, I., {Bartelmann}, M., {Ben{\'{\i}}tez}, N., {Biviano}, A.,
  {Bouwens}, R., {Bradley}, L., {Broadhurst}, T., {Coe}, D., {Czakon}, N., {De
  Petris}, M., {Ford}, H., {Giocoli}, C., {Gottl{\"o}ber}, S., {Grillo}, C.,
  {Infante}, L., {Jouvel}, S., {Kelson}, D., {Koekemoer}, A., {Lahav}, O.,
  {Lemze}, D., {Medezinski}, E., {Melchior}, P., {Mercurio}, A., {Molino}, A.,
  {Moscardini}, L., {Monna}, A., {Moustakas}, J., {Moustakas}, L.~A., {Nonino},
  M., {Rhodes}, J., {Rosati}, P., {Sayers}, J., {Seitz}, S., {Zheng}, W.,
  {Zitrin}, A., Dec. 2014. {The MUSIC of CLASH: Predictions on the
  Concentration-Mass Relation}. apj 797, 34.

\bibitem[{{Merten} et~al.(2015){Merten}, {Meneghetti}, {Postman}, {Umetsu},
  {Zitrin}, {Medezinski}, {Nonino}, {Koekemoer}, {Melchior}, {Gruen},
  {Moustakas}, {Bartelmann}, {Host}, {Donahue}, {Coe}, {Molino}, {Jouvel},
  {Monna}, {Seitz}, {Czakon}, {Lemze}, {Sayers}, {Balestra}, {Rosati},
  {Ben{\'{\i}}tez}, {Biviano}, {Bouwens}, {Bradley}, {Broadhurst}, {Carrasco},
  {Ford}, {Grillo}, {Infante}, {Kelson}, {Lahav}, {Massey}, {Moustakas},
  {Rasia}, {Rhodes}, {Vega}, and {Zheng}}]{2015ApJ...806....4M}
{Merten}, J., {Meneghetti}, M., {Postman}, M., {Umetsu}, K., {Zitrin}, A.,
  {Medezinski}, E., {Nonino}, M., {Koekemoer}, A., {Melchior}, P., {Gruen}, D.,
  {Moustakas}, L.~A., {Bartelmann}, M., {Host}, O., {Donahue}, M., {Coe}, D.,
  {Molino}, A., {Jouvel}, S., {Monna}, A., {Seitz}, S., {Czakon}, N., {Lemze},
  D., {Sayers}, J., {Balestra}, I., {Rosati}, P., {Ben{\'{\i}}tez}, N.,
  {Biviano}, A., {Bouwens}, R., {Bradley}, L., {Broadhurst}, T., {Carrasco},
  M., {Ford}, H., {Grillo}, C., {Infante}, L., {Kelson}, D., {Lahav}, O.,
  {Massey}, R., {Moustakas}, J., {Rasia}, E., {Rhodes}, J., {Vega}, J.,
  {Zheng}, W., Jun. 2015. {CLASH: The Concentration-Mass Relation of Galaxy
  Clusters}. apj 806, 4.

\bibitem[{{Molin{\'e}} et~al.(2017){Molin{\'e}}, {S{\'a}nchez-Conde},
  {Palomares-Ruiz}, and {Prada}}]{2017MNRAS.466.4974M}
{Molin{\'e}}, {\'A}., {S{\'a}nchez-Conde}, M.~A., {Palomares-Ruiz}, S.,
  {Prada}, F., Apr. 2017. {Characterization of subhalo structural properties
  and implications for dark matter annihilation signals}. mnras 466,
  4974--4990.

\bibitem[{{Mu{\~n}oz-Cuartas} et~al.(2011){Mu{\~n}oz-Cuartas}, {Macci{\`o}},
  {Gottl{\"o}ber}, and {Dutton}}]{2011MNRAS.411..584M}
{Mu{\~n}oz-Cuartas}, J.~C., {Macci{\`o}}, A.~V., {Gottl{\"o}ber}, S., {Dutton},
  A.~A., Feb. 2011. {The redshift evolution of {$\Lambda$} cold dark matter
  halo parameters: concentration, spin and shape}. mnras 411, 584--594.

\bibitem[{{Navarro} et~al.(1996){Navarro}, {Frenk}, and
  {White}}]{1996ApJ...462..563N}
{Navarro}, J.~F., {Frenk}, C.~S., {White}, S.~D.~M., May 1996. {The Structure
  of Cold Dark Matter Halos}. apj 462, 563.

\bibitem[{Navarro et~al.(1997)Navarro, Frenk, and White}]{Navarro:1996gj}
Navarro, J.~F., Frenk, C.~S., White, S. D.~M., 1997. {A Universal density
  profile from hierarchical clustering}. Astrophys. J. 490, 493--508.

\bibitem[{{Navarro} et~al.(2004){Navarro}, {Hayashi}, {Power}, {Jenkins},
  {Frenk}, {White}, {Springel}, {Stadel}, and {Quinn}}]{2004MNRAS.349.1039N}
{Navarro}, J.~F., {Hayashi}, E., {Power}, C., {Jenkins}, A.~R., {Frenk}, C.~S.,
  {White}, S.~D.~M., {Springel}, V., {Stadel}, J., {Quinn}, T.~R., Apr. 2004.
  {The inner structure of {$\Lambda$}CDM haloes - III. Universality and
  asymptotic slopes}. mnras 349, 1039--1051.

\bibitem[{{Neto} et~al.(2007){Neto}, {Gao}, {Bett}, {Cole}, {Navarro}, {Frenk},
  {White}, {Springel}, and {Jenkins}}]{2007MNRAS.381.1450N}
{Neto}, A.~F., {Gao}, L., {Bett}, P., {Cole}, S., {Navarro}, J.~F., {Frenk},
  C.~S., {White}, S.~D.~M., {Springel}, V., {Jenkins}, A., Nov. 2007. {The
  statistics of {$\Lambda$} CDM halo concentrations}. mnras 381, 1450--1462.

\bibitem[{{Okabe} et~al.(2013){Okabe}, {Smith}, {Umetsu}, {Takada}, and
  {Futamase}}]{2013ApJ...769L..35O}
{Okabe}, N., {Smith}, G.~P., {Umetsu}, K., {Takada}, M., {Futamase}, T., Jun.
  2013. {LoCuSS: The Mass Density Profile of Massive Galaxy Clusters at z =
  0.2}. apjl 769, L35.

\bibitem[{{Okoli} and {Afshordi}(2016)}]{2016MNRAS.456.3068O}
{Okoli}, C., {Afshordi}, N., Mar. 2016. {Concentration, ellipsoidal collapse,
  and the densest dark matter haloes}. mnras 456, 3068--3078.

\bibitem[{{Prada} et~al.(2012){Prada}, {Klypin}, {Cuesta}, {Betancort-Rijo},
  and {Primack}}]{2012MNRAS.423.3018P}
{Prada}, F., {Klypin}, A.~A., {Cuesta}, A.~J., {Betancort-Rijo}, J.~E.,
  {Primack}, J., Jul. 2012. {Halo concentrations in the standard {$\Lambda$}
  cold dark matter cosmology}. mnras 423, 3018--3030.

\bibitem[{{Rasia} et~al.(2013){Rasia}, {Borgani}, {Ettori}, {Mazzotta}, and
  {Meneghetti}}]{2013ApJ...776...39R}
{Rasia}, E., {Borgani}, S., {Ettori}, S., {Mazzotta}, P., {Meneghetti}, M.,
  Oct. 2013. {On the Discrepancy between Theoretical and X-Ray
  Concentration-Mass Relations for Galaxy Clusters}. apj 776, 39.

\bibitem[{{Rasia} et~al.(2012){Rasia}, {Meneghetti}, {Martino}, {Borgani},
  {Bonafede}, {Dolag}, {Ettori}, {Fabjan}, {Giocoli}, {Mazzotta}, {Merten},
  {Radovich}, and {Tornatore}}]{2012NJPh...14e5018R}
{Rasia}, E., {Meneghetti}, M., {Martino}, R., {Borgani}, S., {Bonafede}, A.,
  {Dolag}, K., {Ettori}, S., {Fabjan}, D., {Giocoli}, C., {Mazzotta}, P.,
  {Merten}, J., {Radovich}, M., {Tornatore}, L., May 2012. {Lensing and x-ray
  mass estimates of clusters (simulations)}. New Journal of Physics 14~(5),
  055018.

\bibitem[{{Sereno} and {Covone}(2013)}]{2013MNRAS.434..878S}
{Sereno}, M., {Covone}, G., Sep. 2013. {The mass-concentration relation in
  massive galaxy clusters at redshift ${\sim}$ 1}. mnras 434, 878--887.

\bibitem[{{Sereno} and {Zitrin}(2012)}]{2012MNRAS.419.3280S}
{Sereno}, M., {Zitrin}, A., Feb. 2012. {Triaxial strong-lensing analysis of the
  z > 0.5 MACS clusters: the mass-concentration relation}. mnras 419,
  3280--3291.

\bibitem[{{Sheth} et~al.(2001){Sheth}, {Mo}, and
  {Tormen}}]{2001MNRAS.323....1S}
{Sheth}, R.~K., {Mo}, H.~J., {Tormen}, G., May 2001. {Ellipsoidal collapse and
  an improved model for the number and spatial distribution of dark matter
  haloes}. mnras 323, 1--12.

\bibitem[{{Springel}(2005)}]{2005MNRAS.364.1105S}
{Springel}, V., Dec. 2005. {The cosmological simulation code GADGET-2}. mnras
  364, 1105--1134.

\bibitem[{Wechsler et~al.(2002)Wechsler, Bullock, Primack, Kravtsov, and
  Dekel}]{0004-637X-568-1-52}
Wechsler, R.~H., Bullock, J.~S., Primack, J.~R., Kravtsov, A.~V., Dekel, A.,
  2002. Concentrations of dark halos from their assembly histories. The
  Astrophysical Journal 568~(1), 52.
\newline\urlprefix\url{http://stacks.iop.org/0004-637X/568/i=1/a=52}

\bibitem[{{Wechsler} et~al.(2006){Wechsler}, {Zentner}, {Bullock}, {Kravtsov},
  and {Allgood}}]{2006ApJ...652...71W}
{Wechsler}, R.~H., {Zentner}, A.~R., {Bullock}, J.~S., {Kravtsov}, A.~V.,
  {Allgood}, B., Nov. 2006. {The Dependence of Halo Clustering on Halo
  Formation History, Concentration, and Occupation}. apj 652, 71--84.

\bibitem[{{Wojtak} and {{\L}okas}(2010)}]{2010MNRAS.408.2442W}
{Wojtak}, R., {{\L}okas}, E.~L., Nov. 2010. {Mass profiles and galaxy orbits in
  nearby galaxy clusters from the analysis of the projected phase space}. mnras
  408, 2442--2456.

\bibitem[{{Wojtak} and {Mamon}(2013)}]{2013MNRAS.428.2407W}
{Wojtak}, R., {Mamon}, G.~A., Jan. 2013. {Physical properties underlying
  observed kinematics of satellite galaxies}. mnras 428, 2407--2417.

\bibitem[{{Wu} and {Xue}(2000)}]{2000ApJ...529L...5W}
{Wu}, X.-P., {Xue}, Y.-J., Jan. 2000. {Correlation between the Halo
  Concentration C and the Virial Mass M$_{VIR}$ Determined from X-Ray
  Clusters}. apjl 529, L5--L7.

\bibitem[{{Zhao} et~al.(2003{\natexlab{a}}){Zhao}, {Jing}, {Mo}, and
  {B{\"o}rner}}]{2003ApJ...597L...9Z}
{Zhao}, D.~H., {Jing}, Y.~P., {Mo}, H.~J., {B{\"o}rner}, G., Nov.
  2003{\natexlab{a}}. {Mass and Redshift Dependence of Dark Halo Structure}.
  apjl 597, L9--L12.

\bibitem[{Zhao et~al.(2009)Zhao, Jing, Mo, and Börner}]{0004-637X-707-1-354}
Zhao, D.~H., Jing, Y.~P., Mo, H.~J., Börner, G., 2009. Accurate universal
  models for the mass accretion histories and concentrations of dark matter
  halos. The Astrophysical Journal 707~(1), 354.
\newline\urlprefix\url{http://stacks.iop.org/0004-637X/707/i=1/a=354}

\bibitem[{{Zhao} et~al.(2003{\natexlab{b}}){Zhao}, {Mo}, {Jing}, and
  {B{\"o}rner}}]{2003MNRAS.339...12Z}
{Zhao}, D.~H., {Mo}, H.~J., {Jing}, Y.~P., {B{\"o}rner}, G., Feb.
  2003{\natexlab{b}}. {The growth and structure of dark matter haloes}. mnras
  339, 12--24.

\end{thebibliography}

\end{document}